%% file: egpaper_final.tex
\documentclass[10pt,twocolumn,letterpaper]{article}

\usepackage{iccv}
\usepackage{times}
\usepackage{epsfig}
\usepackage{graphicx}
\usepackage{amsmath}
\usepackage{amssymb}
\usepackage{authblk}
\usepackage{booktabs}
\usepackage{cuted}
\usepackage{capt-of}
\usepackage[breaklinks=true,bookmarks=false]{hyperref}

\iccvfinalcopy % *** Uncomment this line for the final submission

\def\iccvPaperID{23} % *** Enter the ICCV Paper ID here

\ificcvfinal\fi

\begin{document}

%%%%%%%%% TITLE
\title{CGC-Net: Cell Graph Convolutional Network for Grading of Colorectal Cancer Histology Images\vspace{-10pt}}

\author{Yanning Zhou$^{1}$\thanks{This work was conducted while the first author was visiting the Tissue Image Analytics (TIA) Lab at the University of Warwick.} \quad Simon Graham$^{2}$ \quad Navid Alemi Koohbanani$^{2}$ \quad Muhammad Shaban$^{2}$ \\

\vspace{-10pt}\quad Pheng-Ann Heng$^{1}$ \quad Nasir Rajpoot$^{2}$    \\
{\normalsize$^{1}$Department of Computer Science and Engineering, The Chinese University of Hong Kong, China}\\
\vspace{-2pt}{\tt\small \{ynzhou,pheng\}@cse.cuhk.edu.hk}\\
{\normalsize$^{2}$TIA Lab, Department of Computer Science, University of Warwick, UK}\\
\vspace{-2pt}{\tt\small\{s.graham.1,n.alemi-koohbanani,m.shaban,n.m.rajpoot\}@warwick.ac.uk}\\
% %2nd choice
% \leftline{\quad\quad\quad$^{1}$The Chinese University of Hong Kong \quad\quad\quad\quad\quad\quad$^{2}$The University of Warwick}\\
% \leftline{{\tt\small ynzhou@cse.cuhk.edu.hk, s.graham.1@warwick.ac.uk}}\\


}

\maketitle

\ificcvfinal\thispagestyle{empty}\fi
\begin{strip}\centering
\vspace{-60pt}
\includegraphics[width=1.0\textwidth]{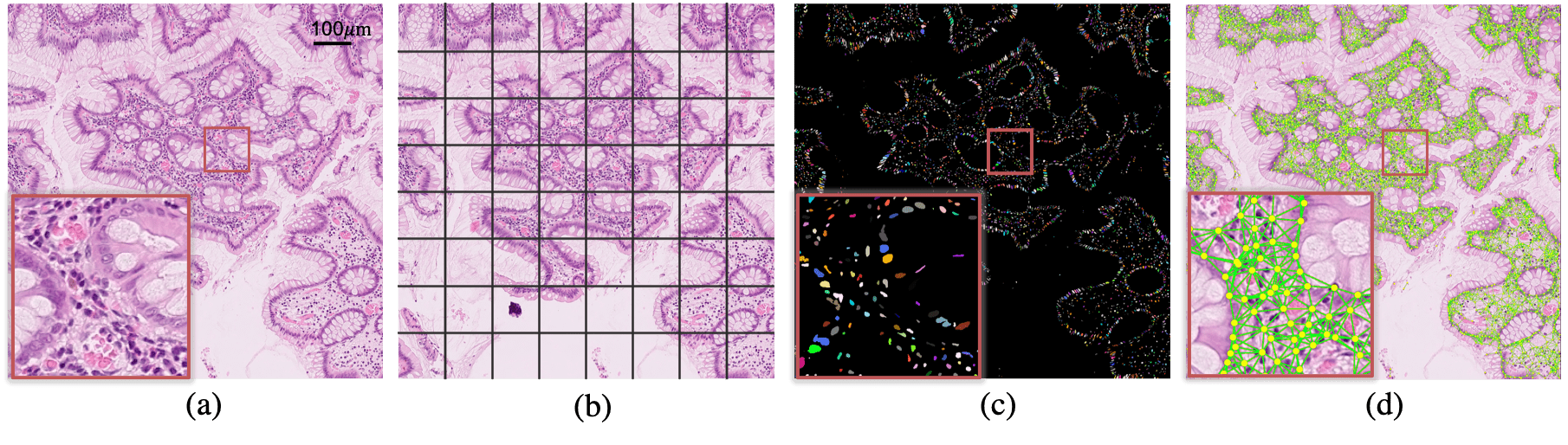}
\captionof{figure}{A histology image (a) is typically broken into small image patches (b) for cancer grading. We propose to utilise the cell graph (d) that is built from individual nuclei after segmentation (c) to model the entire tissue micro-environment for cancer grading.
\label{fig:TOP}}
\end{strip}

%%%%%%%%% ABSTRACT
\begin{abstract}
   Colorectal cancer (CRC) grading is typically carried out by assessing the degree of gland formation within histology images. To do this, it is important to consider the overall tissue micro-environment by assessing the cell-level information along with the morphology of the gland. However, current automated methods for CRC grading typically utilise small image patches and therefore fail to incorporate the entire tissue micro-architecture for grading purposes. To overcome the challenges of CRC grading, we present a novel cell-graph convolutional neural network (CGC-Net) that converts each large histology image into a graph, where each node is represented by a nucleus within the original image and cellular interactions are denoted as edges between these nodes according to node similarity. The CGC-Net utilises nuclear appearance features in addition to the spatial location of nodes to further boost the performance of the algorithm. To enable nodes to fuse multi-scale information, we introduce \textit{Adaptive GraphSage}, which is a graph convolution technique that combines multi-level features in a data-driven way. Furthermore, to deal with redundancy in the graph, we propose a sampling technique that removes nodes in areas of dense nuclear activity. We show that modeling the image as a graph enables us to effectively consider a much larger image (around 16$\times$ larger) than traditional patch-based approaches and model the complex structure of the tissue micro-environment.
   We construct cell graphs with an average of over 3,000 nodes on a large CRC histology image dataset and report state-of-the-art results as compared to recent patch-based as well as contextual patch-based techniques, demonstrating the effectiveness of our method.
\end{abstract}

%%%%%%%%% BODY TEXT
\section{Introduction}
\input{introduction.tex}

\section{Related Work}
\input{relatedwork.tex}

\section{Method}
\input{method.tex}

% \vspace{-10pt}
\section{Experiment}

\input{experiment.tex}

 \section{Conclusion}
\input{conclusion.tex}

\vspace{-5pt}
\section*{Acknowledgments}
{
% NMR was supported by the UK Medical Research Council grant\# MR/P015476/1. The authors are also grateful to the Warwick Global Partnership Fund (GPF) for funding the collaboration between Warwick and CUHK.

% This work was supported in part by the Hong Kong Research Grants Council under General Research Fund (Project No. 14225616).

This work was supported in part by the UK Medical Research Council grant\# MR/P015476/1 and the Hong Kong Research Grants Council under General Research Fund (Project No. 14225616).
The authors are also grateful to the Warwick Global Partnership Fund (GPF) for funding the collaboration between Warwick and CUHK.
}{\small
\bibliographystyle{ieee_fullname}
\bibliography{egbib}
}

\end{document}

%% file: introduction.tex
% background of colon cancer
% importance
\begin{figure*}[t]
    \centering
    \includegraphics[width=1.0\textwidth]{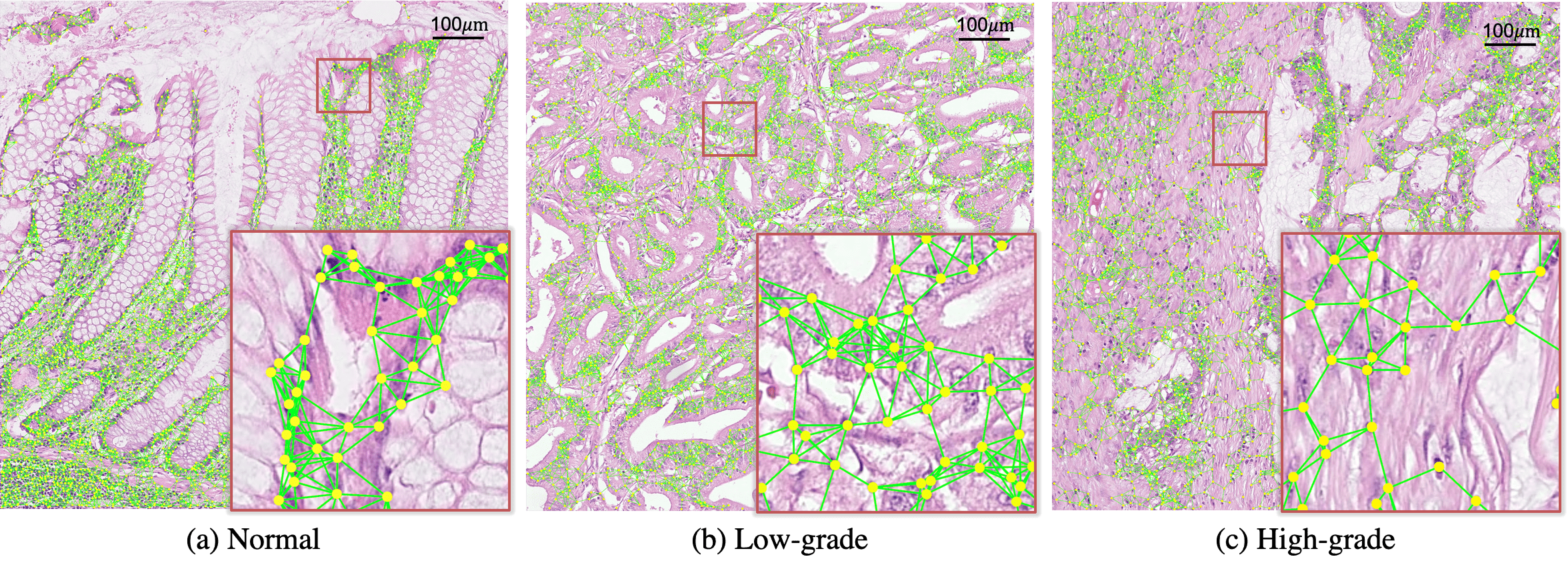}
    \caption{Typical cell graphs from (a) normal, (b) low-grade and (c) high grade images. The blue lines represent the edges and the green dots represent the nuclei (graph nodes). }
    \label{fig:large-graph}
    \vspace{-10pt}
\end{figure*}

Colorectal cancer (CRC) is one of the most common cancers worldwide.
According to the Global Cancer Statistics 2018, CRC is the third most commonly occurring cancer among men and women and is the second most common cause of cancer related mortality  ~\cite{bray2018global}.
% the overall rank of incidence for colorectal is the third among cancers for both men and women while the second in terms of mortality~\cite{bray2018global}.
Among CRC cases, more than $90\%$ of them are colorectal adenocarcinomas (CRA). Based on the degree of glandular formation, CRA can be divided into \textit{low-grade} and \textit{high-grade} cancer, whereby \textit{low-grade} CRA contains well/moderately differentiated adenocarcinomas and \textit{high-grade} CRA contains poorly differentiated/undifferentiated adenocarcinomas~\cite{hamilton2000pathology}.
Grading CRA is a crucial task due to its fundamental role in deciding on an appropriate follow-up treatment and is also indicative of overall patient outcome~\cite{compton2000prognostic}.
% However, it still remains the inter-observer variability for grading.

There has been a recent surge in interest for digital pathology, where tissue samples are digitised with a scanner to create whole slide images (WSIs), enabling efficient storage and management of the specimens. WSIs are stored in a multi-resolution format, where at the highest resolution they can be up to 150,000$\times$100,000 pixels in size and contain hundreds of thousands of cells. The rise of digital pathology has led to the development of computational techniques for automatic quantification and assessment of the tissue, helping reduce the inter-observer variability between pathologists. Furthermore, digital signatures within the tissue can be used to assist with cancer diagnosis and to enable prediction of cancer prognosis and clinical outcome, providing motivation for the use of computational pathology within routine clinical practice.
Recently, several automatic methods have been proposed to grade or classify different cancers including breast, colon and lung cancer~\cite{araujo2017classification,coudray2018classification, graham2018classification}.
 To cope with the very large size of WSIs, the general framework of these methods consists of two steps: patch-based image classification followed by aggregation of patch based classification at the slide level.
First, the WSI is divided into small image patches, where each patch is processed independently. Then, all predictions are combined to obtain the final decision.
% \begin{figure}[t]
%     \centering
%     \includegraphics[width=0.5\textwidth]{latex/compare-framework.png}
%     \caption{Comparison of SOTA context-learning method~\cite{Shaban2019} and our proposed method. }
%     \label{fig:compare-framework}
% \end{figure}

There are two main drawbacks faced by patch based approaches:
First, there is an inherent trade-off between the resolution of each image patch and the context provided. The favourable size for each image patch is data-dependent. For example, in CRA, the grade of cancer is determined by assessing the degree of glandular formation in the tumour. However, the variation in glandular morphology and size leads to a difficulty in defining an appropriate image patch size.
Within Figure~\ref{fig:large-graph}, (a) shows a normal case where the glands have a clear tubular structure. On the other hand, (b) and (c) show cancerous cases where typical glandular appearance is less evident.
Given a set resolution, the maximum image size that can be used is limited by the memory of the GPU. An alternate strategy is to use a lower resolution, which enables the same size patch to provide more context, but at the same time will lose cell-level information that may be diagnostically important.
Second, due to the biases present within the features learned by a convolutional neural network (CNN), features extracted from each image patch may lack an interpretable correspondence to the tissue morphology and glandular structure.

A way to address the above drawbacks is to model nuclear features along with their cellular interactions in the form of a graph, which accounts for both cell-level information and the overal tissue micro-architecture. A graph constructed from cells within a WSI contains many millions of nodes and edges and therefore contains diagnostically important information relating to the tissue micro-architecture that may not be visible by manual inspection. Recently, cell graphs using graph theory have been studied , in order to capture the functional organisation of cells~\cite{yener2016cell}.
% The spatial distributions and structural patterns of the cell communities are associated with the specific functional they perform.
Previous work computed predefined graph-based features for cancer diagnosis~\cite{demir2005augmented,bilgin2007cell,bilgin2010ecm, sirinukunwattana2018novel,javed2018cellular}.
However, it remains unclear how to select and combine graph-level features to best represent the complex organisation of cells in different tissue components.

In this paper, we propose a novel and general framework called Cell Graph Convolutional Network (CGC-Net) for histology image classification, based on the recent development of graph convolutional neural networks, and demonstrate its effectiveness for grading of colorectal cancer histology images.
A cell graph is directly constructed from an image, where the nuclei are regarded as the nodes and the potential cellular interactions as edges of the graph. To acquire accurate node features, we apply a nuclear segmentation network and extract appearance features based on the segmented foreground instances.
Instead of computing predefined graph-level features from the cell graph, the proposed CGC-Net takes the entire graph as input to obtain a compact representation of the tissue micro-environment for cancer grading in an end-to-end manner.
Within our proposed model, we introduce the \textit{Adaptive GraphSage} as a new graph convolution module to enforce the fusion of multi-level node features in a data-driven manner, followed by the graph clustering module to coarsen the graph. We then employ the learned hierarchical features for graph-based classification.
We would like to emphasize that it is non-trivial to construct a cell graph suitable for the graph convolutional network due to the large number of cells in a histology image.
Therefore, a representative nuclei sampling strategy is proposed to reduce the number of nodes and edges according to the relative inter-node distance.
% The selected nuclei are established edges by their relevant spatial distance.

Overall, our main contributions can be summarized as follows:
\begin{itemize}
    \item The CGC-Net is the first network  of its kind for cancer grading that bridges the gap between the deep learning framework and the conventional cell graph.

    \item A general cell graph construction pipeline with a representative nuclei sampling strategy that utilizes nuclear appearance and spatial information.
    % The representative nuclei sampling strategy effectively reduces the graph size, whilst preserving the tissue micro-environment structure.
    \item The CGC-Net utilizes the nuclei rather than small patches as descriptors, where cluster visualization leads to better biological insight and interpretability.
    % The visualization of each cluster demonstrates that the network can effectively learn the local patterns and assign them into different clusters.
    \item A comprehensive study on a large colorectal cancer dataset. Results show the proposed CGC-Net outperforms other state-of-the-art methods.
    % Additionally, ablation studies show the effectiveness of the graph construction framework as well as the network structure.

\end{itemize}

% However, the predefined graph-based features are not powerful enough to represent the complex morphometric in different tissue components.

% It hypothesizes the cellular communication orderly organizes them into tissues to perform a specific function.
% learn the different structure components in the image by modeling the cell communication~\cite{yener2016cell,sirinukunwattana2018novel}.
% Nodes of a cell-graph are associated with individual cells and the edge denotes the potential communication between them.
% The type and spatial distribution of these different tissues can be used to recognize the progression of tumours.
% aims to encode the pairwise relation between connected cell by statistically
% assigning a link between them

% deep learning success
% classification patch

% on the other hand, cell graph also been study
% but

% we first attempt to bridge the gap, xxx

%% file: relatedwork.tex
\noindent\textbf{Cancer grading in histology images:}
% Histological images are usually larger than the natural image to capture both cell level and structure level visual cues for disease analysis and diagnosis.
In the literature, earlier methods relied on different hand-crafted feature extraction techniques, including nuclear appearance features (e.g. colour, texture and shape)~\cite{diamond2004use,dundar2011computerized} and morphological features~\cite{zana2001segmentation,nguyen2012prostate} to distinguish between different grades of cancer. 
Recently, deep learning methods have been widely used in various cancer grading tasks for a variety of tissues, including: lung; breast and colorectal cancer~\cite{coudray2018classification,spanhol2016breast,awan2017glandular}. Typically, the grading framework utilises a CNN for patch-level classification and then individual image patch predictions are combined to yield the overall result.  
Different strategies have been proposed to utilise contextual information to obtain a better prediction. for example, ~\cite{gecer2018detection,liu2017detecting} combined multi-resolution information in either image space or feature space, whereas~\cite{htadaptive2018} used an adaptive patch selection approach.
In~\cite{wsicontext2018}, the authors conducted a comprehensive study on multi-scale information fusion methods and proved that utilizing LSTM units to embed features with larger context results in a superior performance.

In the case of colorectal adnocarcinoma, Awan \textit{et al.}~\cite{awan2017glandular} proposed a novel \textit{Best Alignment Metric} (BAM) to measure glandular morphology for classification, highlighting the importance of the glands within colorectal cancer grading. As a prerequisite, the BAM metric relies on a good quality gland segmentation, which has been successfully explored in recent work~\cite{graham2019mild}.
Recently, Shaban \textit{et al.}~\cite{Shaban2019} proposed a context-aware network which uses an attention mechanism to aggregate information from larger contextual regions and outperforms other context-based methods as well as domain-oriented methods. 
However, despite some of the aforementioned methods incorporating additional context, they are still limited by a pre-defined patch size and do not necessarily incorporate the entire tissue micro-environment. 
% The common procedure be divided into two steps: patch-based feature and image-level result aggregation.

\noindent\textbf{Cell graph:}
Cell graphs aim to model the relationship between different cells and the tissue micro-environment utilizing graph features~\cite{schnorrenberg1996computer, weyn1999computer, demir2005augmented}. 
Within a cell graph, the nuclei or cell clusters are regarded as vertices and the potential signal between them are regarded as the edges.
Based on the assumption that adjacent cells are more likely to interact, the graph can be constructed via Delaunay triangulation \cite{keenan2000automated} or the K-nearest-neighbour method ~\cite{bilgin2007cell}.
After cell graph construction, the distribution of cell level features are converted into global features and combined with other predefined graph features to train a machine learning algorithm, e.g. SVM, Bayesian, and KNN~\cite{oztan2012follicular}.  
Bilgin \textit{et al.}~\cite{bilgin2010ecm}, proposed the ECM-aware cell graph for bone tissue modeling and classification by incorporating colour information and assigning a colour label for each node.
Aside from the task of cancer diagnosis, Sirinukunwattana \textit{et al.} ~\cite{sirinukunwattana2018novel} leveraged of the cell-cell interaction between different cell types as tissue phenotypic signatures and used an unsupervised learning approach to group the different tissue types for distant metastasis estimation.
All the above mentioned methods need to define and extract graph-level features for further classification or clustering.
% Recently, Shi et al.

\noindent\textbf{Graph neural network:}
Earlier Graph neural network methods~\cite{gori2005, micheli2009neural,Scarselli2009TheGN} utilise recurrent neural networks for neighbor information propagation.
To reduce the expensive computation, recent work utilised the concept of convolution and proposed various Graph Convolutional Networks (GCNs), which can be divided into spectral-based GCN ~\cite{bruna2014,NIPS2016_6081,kipf2017} and spatial-based GCN~\cite{niepert2016learning,hamilton2017inductive}.
Kipf and Welling ~\cite{kipf2017} proposed a localised first-order approximation of spectral graph convolutions for scalable semi-supervised learning. 
GraphSage was proposed in~\cite{hamilton2017inductive}, which introduces aggregation functions for message parsing and a batch-training strategy to improve the scalability of large graphs.
To learn hierarchical features for better graph-level classification and to reduce the computational complexity, different graph pooling methods have been proposed to reduce the graph size~\cite{zhang2018end,gunet,ying2018hierarchical}. 
Ying et al.~\cite{ying2018hierarchical} proposed the differential graph pooling method which utilises another graph convolution layer to generate the assignment matrix for each node.
Xu et al.~\cite{jknet} proposed to leverage different neighborhood ranges adaptively for better feature representation. 
Recently,~\cite{gcncervical} attempted to model the relation-aware representation for cervical cell classification.
However, no work has been done to utilise GCNs to model the tissue micro-environment.

%% file: method.tex
% \subsection{Motiviation}

% The analysis of shape and appearance features of individual nuclei within histology images is of high importance because they can reveal critical predictive markers for diagnosing cancer [ref]. However, analyzing individual nuclei does not take into account the various nuclear communities that exist within the tissue micro-environment which can provide important information regarding patient outcome.

% Effectively combining cell-level features and the interactions between communities of cells remains a key challenge.

% Analyzing small image patches for cancer detection does not take into account the contextual information, that pathologists often rely on to make a diagnosis. Hence, onsidering both cellular and topological information, in this paper, we attempt to construct a cell graph for each image.
% Our motivations for using graph are:
% \textit{Interpretability}: Using nuclei descriptor and their potential interactions which can give more insight about tumor micro-environment
% \textit{Extensibility}: Considering essential information of tumor micro-environment by applying  cellular and topological information in the node and edge of the graph, respectively
% \textit{Flexibility}: By parsing message through edges, we can adaptively capture lesions with different scales and does not necessarily require pre-defined patch size.

\subsection{Graph Convolutional Network}

Different from the conventional convolution which operates on the regular grid in the Euclidean space, the graph convolution extends the information aggregation to the non-Euclidean space to allow incorporation of irregular data structure.

A graph is defined as $G = (V, E)$, which consists of a node set $V$ with d-dimensional node features $x_{i} \in \mathbb{R}^{d}$ for $i \in V$ and edge set $E$, where $e_{i,j} = (i, j)\in E$ denotes an edge.
An adjacency matrix $A\in \mathbb{R}^{n \times n}$ has non-zero entry $A_{ij} > 0$ if $e_{ij} \in E$.
Let $h^{(l)}_{i} \in \mathbb{R}^ d$ denote the hidden features in the $l$-th layer for node $i$, then its analogous neighborhood is $\widetilde{N}(i)=\left\{i\right\}\cup\left\{ j \in V| e_{ij}>0\right\}$.
Here, we use $h^{(0)}_{i} = x_{i}$ for the input layer.
A typical graph convolution operation can be written as:
\begin{equation}
% h^{(l)}_{i} = f\left ({h^{(l-1)}_{i}, }, ;\theta^{(l)} \right ),
% f\left\(\left\{h^{(l-1)}_{j}, \forall j \in\tilde{N}(i)\right\}, \theta^{(l)}\right\)
 h^{(l)}_{i} = \sigma \left( W^{(l)} \cdot Agg\left\{h^{(l-1)}_{j}, \forall j \in{\widetilde{N}}(i)\right\} \right),
 \label{eq:agg}
\end{equation}
where $Agg\left\{ \cdot \right\}$ is a pre-defined aggregation function, $W^{(l)}$ is the learnable weight in the $l$-th layer shared by all the nodes and $\sigma$ is the non-linear function, where specifically we use ReLU in our experiments.
% It takes the hidden features from node itself and its neighbors as inputs to aggregate information.

\subsection{Cell Graph Construction}
\label{sec:graph-construct}
% For each image, a cell graph is constructed based on the cell segmentation results.
% \begin{figure*}[t]
%     \centering
%     \includegraphics[width=1.0\textwidth]{latex/graph-construct.png}
%     \caption{Graph construction pipeline. The original image(a) is passed through a nuclei segmentation network to get the nuclei mask(b), which is further used to extract nuclei features for the cell graph. The representative nuclei sampling strategy is used to reduce the size of the cell graph from (c) to (d). }
%     \label{fig:large-graph}
% \end{figure*}

Constructing a meaningful graph which reflects the potential interactions between cells is a vital part of our analysis.
Each image is converted to a cell graph, where nodes are nuclear descriptors and edges are the potential interactions between cells.
In order to construct the graph, we complete the following steps: \romannumeral 1) nuclear instance segmentation to extract node features; \romannumeral 2) representative node sampling to remove redundancy in the graph and \romannumeral 3) graph edge configuration to define potential cellular interactions. \\
\textbf{Nuclear instance segmentation:}
Precise nuclear instance segmentation leads to more reliable node features in the cell graph. Therefore, due to its high performance, we use CIA-Net~\cite{zhou2019cia} to accurately delineate the boundaries of each nucleus. CIA-net is a contour-aware network with two interacting branches that aggregate both contour and nuclear features for a superior result.

\noindent\textbf{Cell nuclei feature extractor:}
The nuclear masks obtained via CIA-Net are used to extract nuclear shape and appearance features to strengthen the diagnostic capability of our graph-based approach for colorectal cancer grading. In order to select the most predictive nuclear descriptors, we first implemented a random forest model to classify nuclei as either epithelial, inflammatory or spindle-shaped and then utilised feature selection to choose the 16 most predictive features. We chose to select features that were predictive of the nuclear category because the features can subsequently help to indirectly encode the category of each nucleus. Here, we chose to keep the raw descriptors rather than the predicted categories because we assume that they may be more informative for predicting the grade of cancer.
% We also calculate the centroid of the mask as the location features for that cell.
In addition, we incorporate the centroid coordinates and therefore in total use seventeen nuclear descriptors: mean nuclei intensity; average fore-/background difference; standard deviation of nuclei intensity; skewness of nuclei intensity; mean entropy of nuclei intensity; GLCM of dissimilarity; GLCM of homogeneity; GLCM of energy; GLCM of ASM; eccentricity; area, maximum length of axis; minimum length of axis; perimeter, solidity; orientation and centroid coordinates.

\begin{figure*}[th]
    \centering
    \includegraphics[width=1.05\textwidth]{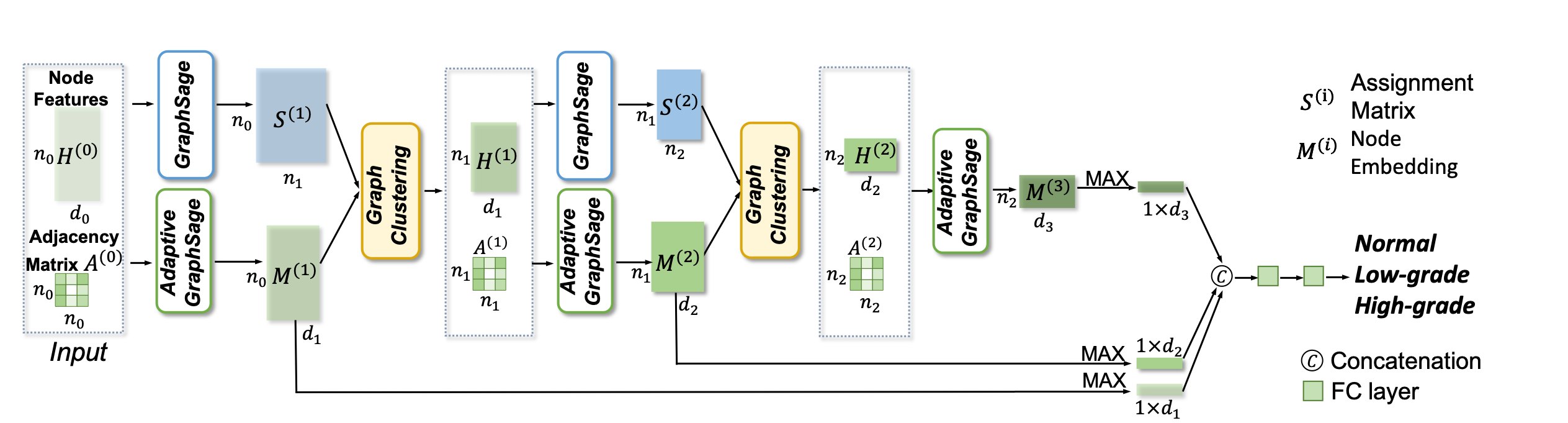}
    \caption{Overview of the CGC-Net.}
    \label{fig:overview}
\vspace{-10pt}
\end{figure*}

\begin{figure}[ht]
    \centering
    \includegraphics[width=0.43\textwidth]{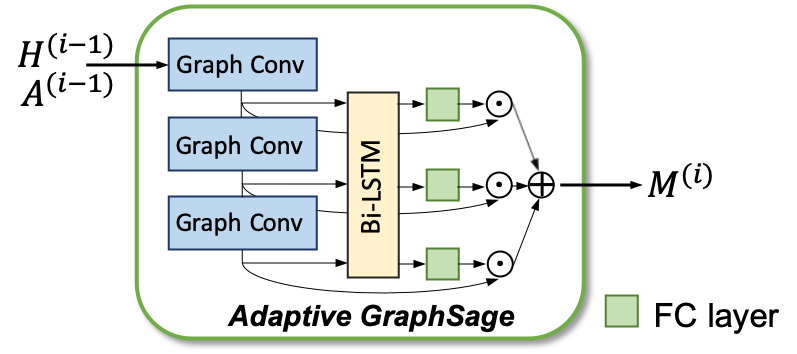}
    \caption{Detailed structure of the Adaptive GraphSage module.}
    \label{fig:ags}
    \vspace{-10pt}
\end{figure}
% Apart from appearance features, we also calculate the centroid of the mask as its location features. These features are concatenated with appearance features to generate the final node features.
\noindent\textbf{Representative nuclei sampling strategy:}
% The original image should not be too small to capture the high-level gland structure information in the cell graph.
% Therefore, most of the images consists of a large amount of cells.
Utilizing all nuclei in the image as nodes within a graph is undesirable because of the following reasons:
First, some regions are dense with many cells containing similar features and therefore it is unnecessary to incorporate them all within the graph.
Second, some graphs contain a huge amount of nuclei and therefore utilizing them all within the graph is very computationally expensive.

To resolve this problem, we propose to sample $a$-ratio of the representative nuclei, instead of using all of them.
Specifically, we use the Farthest Point Sampling (FPS) method ~\cite{eldar1997farthest} to choose a subset of nuclei, where each nucleus has the farthest distance to the selected nuclei collection.
Compared to random sampling, it effectively alleviates the problem of removing nuclei within sparse areas.
Furthermore, to prevent over-fitting we sample $b$-ratio $(b<a)$ of nuclei randomly and add them to the selected subset. We choose $a=0.35$ and $b=0.15$ in all experiments.

\noindent\textbf{Graph edge configuration:}
In the cell graph, we define an edge as the potential interaction between two nuclei.
We hypothesise that the cells with a smaller Euclidean distance are more likely to interact.
To this end, we assign an edge between two nuclei if they are within a fixed distance from each other. Moreover, the maximum degree of each node is set to $k$ corresponding to its $k$-nearest neighbors.
% To avoid the edge cross the lumen and endothelium regions which are the white spaces in the image, we add a maximum distance threshold for the edge.
Formally, the adjacency matrix can be written as follow:
\begin{equation}
    A_{ij}\left\{\begin{matrix}
1  & \text{if} \ j \in KNN(i) \ \text{and}\ D(i,j)<d , \\
0 &  \text{otherwise}.
\end{matrix}\right.
\end{equation}
$D(\cdot, \cdot)$ denotes Euclidean distance.
% \begin{equation}
%     A_{ij}\left\{\begin{matrix}
% 1  & if\ j\in\left\{Rank_{a}(\left\{D(i,a)\right\},k)\right\},   D(i,a)<d  \and\ a\neq i, \\
% 0 &  otherwise.
% \end{matrix}\right
% \end{equation}

% \begin{equation}
%     A_{ij}^{'}\left\{\begin{matrix}
% p/\sum^{n}_{j=1,j\neq i}A_{ij}  & if\ i\neq j \\
%  1-p & if \ i = j
% \end{matrix}\right.,

\subsection{Cell Graph Network Architecture}

% The proposed CG-Net is a stack of graph convolutional layers and graph pooling layers.
After constructing the cell graph, the task of colorectal cancer grading can be considered as a graph-level classification problem.
We proposed the CGC-Net equipped with a stack of graph convolution and graph pooling modules.
The graph convolution aggregates features from the nodes' local neighbors, which in our case is the nuclei along with their interactions.
To enable the nodes to fuse multi-scale features according to the contextual structure adaptively, we proposed Adaptive GraphSage which combines multi-level features in a data-driven way.
After generating the node's hidden embedding, the graph clustering operation coarsens the graph by assigning the nodes to different groups, which can be considered as an extension of the standard pooling operation.

\noindent\textbf{Adaptive GraphSage:} Given node features and edge information, various types of graph convolution can be used to learn the nodes' hidden representation.
In~\cite{hamilton2017inductive}, it processes the predefined aggregation functions $ Agg \left( \cdot \right)$ in Eq.~\ref{eq:agg} including mean, sum and max function.
Then, it combines the multi-level node representation by concatenation and applies the operation $k$ times to capture the $k$-hop neighbors' information.

Although the node embedding after GraphSage contains multi-level neighborhood information, it cannot adaptively assign weights to the features according to the local topological structure around the target node.
In other words, it can only fuse the multi-level embedding features in the same way for all nodes in that graph, which is not suitable for the cell graph because we want to capture the gland structure at various scales.

Inspired by~\cite{jknet}, we propose a learnable pattern to aggregate multi-level embedding features for each node to address this issue.
% In particular, let $h^{(l)}_{v}$, $l=1,2 \dots k$ denote the $l$-hop aggregated feature of node $v$ in $l$-th layer.
In particular, the proposed Adaptive GraphSage stacks $k$ graph convolutions, which means that each node can aggregate information from its $k$-hop neighbors.
% We consider $\left \{ h^{(1)}_{v},h^{(2)}_{v} \dots h^{(k)}_{v} \right \}$ as sequential inputs with dependent information and feed them into a bi-directional LSTM to acquire the forward and backward hidden embeddings $f^{(l)}_{v}$ and $b^{(l)}_{v}$ for each feature.
We consider the intermediate outputs from graph convolutions $\left \{ h^{(1)}_{v},h^{(2)}_{v} \dots h^{(k)}_{v} \right \}$ as sequential inputs with dependent information and feed them into a bi-directional LSTM to acquire the forward and backward hidden embeddings $f^{(l)}_{v}$ and $b^{(l)}_{v}$ for each feature.
Then for each layer $h^{(l)}_{v}$, the concatenation of forward and backward hidden embeddings $\left [ f^{(l)}_{v} | b^{(l)}_{v} \right ]$ are passed through a linear mapping function followed by a Softmax to get the importance score $s^{(l)}_{v}$.
Finally, the representation of each node is obtained by performing a weighted sum of the multi-level features $m_{v} = \sum _{l}s^{(l)}_{v}\cdot h^{(l)}_{v}$.

The proposed Adaptive GraphSage utilises an attention mechanism to enable multi-scale feature fusion.
This mechanism allows the network to generate an effective node representation according to its local structure, adaptively.   \\
% \textb{Graph clustering operation.}
\noindent\textbf{Graph clustering module:}
After the input is passed through the Adaptive GraphSage, the node features contain the local contextual information.
However, the flat structure remains a drawback because the the hierarchical structure is lost when using global mean/max pooling to get the graph-level prediction.
% For cell graph classification, either the local cell patterns (e.g. appearance and density) or the middle level information (e.g. cell community and gland structure) are contributed to the accuracy.
Therefore, the clustering operation is necessary to extract more abstract features for hierarchical representation. We make use of the graph clustering method used by Ying \textit{et al.}~\cite{ying2018hierarchical} that utilises another graph convolution for node assignment prediction in parallel with feature extraction.

To be concise, let $H^{(i)} \in \mathbb{R}^{n_{i} \times d_{i}}$ denote the features for all nodes after $i$-th graph clustering and $A^{(i)} \in \mathbb{R}^{n_{i} \times n_{i}}$ denotes the adjacency matrix ($H^{(0)}$ and $A^{(0)}$ are the input features and adjacency matrix).
One Adaptive GraphSage is applied to generate the embedding matrix $M^{(i)}$.
Meanwhile, the nodes are passed through another GraphSage followed by a linear function to generate the assignment matrix $S^{(i)} \in \mathbb{R}^{n_{i-1} \times n _{i}}$.
$S^{(i)}$ denotes the probability of each node being assigned to each cluster, e.g. $S_{pq}$ denotes the probability of assigning the $p$-th node to the $q$-th cluster.

After we get the $M^{(i)}$ and $S^{(i)}$, the clusters are considered as new nodes for the following layer, where the clusters' features and corresponding adjacency matrix are:
\begin{equation}
    H^{(i)} = S^{(i)^{T}}M^{(i)},
\end{equation}
\begin{equation}
    A^{(i)} = S^{(i)^{T}}A^{(i-1)}S^{(i)}
\end{equation}
% To reduce the parameter, we utilize the standard GraphSage to generate multi-level features.
% Then, they are directly concatenated and pass through a linear function and a Softmax layer to generate the weight of each node been assigned to each cluster Softmax
% For convenient, we rewrite all the nodes' hidden feature in $l$-th layer as $H^{l} \in \mathbb{R}^{n_{l} \times d_{l}}$.
% $S_{n_{i}} \in \mathbb{R}^{n_{i} \times n_{i+1}}$
% \textbf{}
\noindent\textbf{Over-smooth problem:}
The literature reports that there exists the \textit{over-smooth problem} for graph convolution networks~\cite{chen2019multi,li2018deeper}.
To alleviate this problem, we apply the re-weighted scheme that was originally proposed by Chen \textit{et al.}~\cite{chen2019multi}. Concretely, this scheme is defined as:
\begin{equation}
    A_{ij}^{'}\left\{\begin{matrix}
p/\sum^{n}_{j=1,j\neq i}A_{ij}  & \text{if}\ i\neq j \\
 1-p & \text{if} \ i = j
\end{matrix}\right.,
\end{equation}
where $p=0.4$ in our experiment.

\noindent\textbf{Combining Hierarchical features for graph-level classification:}
We utilise a max operation for the node embeddings at each stage to get a fixed-size representation.
Then the concatenation of multi-level representations is fed into the linear layer to get the prediction for 3-class classification.
The whole network is trained with cross-entropy loss.
% as follows:
% \begin{equation}
%     \mathcal{L} = - \sum_{k=1}^{n}\left ( p_{k}\ast \log q_{k} \right ),
% \end{equation}
% where $p_{k} \in \left\{0, 1 \right\} $ is the ground truth label and $q_{k} \in \left [0,1  \right ]$ is the prediction.
% \begin{equation}
%     y_{pred} = Softmax(MLP())
% \end{equation}

\subsection{Implementation}
% We utilized state-of-the-arts nuclei segmentation network~\cite{zhou2019cia,grahamhover} to generate nuclei masks.
All the node features are normalised by subtracting the mean and dividing by its standard deviation channel-wise.
To prevent over-fitting, Dropout~\cite{srivastava2014dropout} is used with $p=0.2$ during training.
The CGC-Net is implemented using PyTorch~\cite{paszke2017automatic} with the geometric deep learning package~\cite{Fey/Lenssen/2019}.
We use Adam optimization with an initial learning rate of $1e^{-3}$. All models are trained for 30 epochs with a batch size of 40.
The learning rate is dropped to $\left\{1e^{-4},1e^{-5}  \right\}$ after 10 and 20 epochs respectively and the weight decay is set to be $1e^{-4}$.
\textcolor{black}{The model has 1.44M parameters. The training process takes 12 hours on a server with 4 NVIDIA TITAN V GPUs. It takes 6 minutes to process 10500 patches($1792\times1792$)} on a single GPU.
% All models are trained with the learning rate$\left\{ 1e^{-3},1e^{-4},1e^{-5}   \right\}$

%% file: experiment.tex
\subsection{Dataset and Evaluation Metrics}
\noindent\textbf{Colorectal Cancer(CRC) dataset:}
% The proposed method is evaluated on the CRC dataset~\cite{awan2017glandular}, which consists of 139 images taken from WSIs from University Hospitals Coventry and Warwickshire with an average size of 4548$\times$7520 at 20$\times$ magnification.
The proposed method is evaluated on the CRC dataset~\cite{awan2017glandular}, which consists of 139 images taken from WSIs with an average size of 4548$\times$7520 at 20$\times$ magnification.
The images are divided into \textit{normal}, \textit{low grade} and \textit{high grade} based on the degree of gland differentiation.
To conduct a fair comparison, we split the dataset the same way as ~\cite{Shaban2019} into three folds for cross-validation.
We extract patches with a size of 1792$\times$1792 pixels for cell graph construction, which is the same size that the context-aware learning method ~\cite{Shaban2019} used.
\textcolor{black}{Similar to~\cite{Shaban2019}, majority voting is used to generate the image-level prediction.}
To evaluate the performance, we use the average accuracy at both image-level (4548$\times$7520) and patch-level (1792$\times$1792).
% , which refers to the average percentage of images being assigned to the correct class across three folds.

\noindent\textbf{Colorectal nuclear segmentation and phenotypes (CoNSeP) dataset:}
To train the nuclear segmentation network for graph construction, we use the CoNSeP dataset~\cite{grahamhover}.
CoNSeP consists of 41 H\&E stained images with 1000$\times$1000 pixels at 40$\times$ magnification extracted from 16 CRA WSIs. Image patches were chosen to be representative of the tissue within CRA histology images and therefore we expect a method trained no this dataset to generalise well to the images used for graph construction.
% CoNSeP consists of 41 H\&E stained images with 1000$\times$1000 pixels at 40$\times$ magnification extracted from 16 CRA WSIs from University Hospitals Coventry and Warwickshire. Image patches were chosen to be representative of the tissue within colorectal adenocarcinoma histology images and therefore we expect a method trained no this dataset to generalize well to the images used for graph construction.
\vspace{-10pt}
\subsection{Experimental Results}

\begin{table*}[t]
    \centering
    \begin{tabular}{c|c|c|c|c}
    \toprule
         Node Features & GC & Sampling& Patch Accuracy & Image Accuracy \\
    \toprule
          Appearance \& Spatial &\textit{GS}& Fuse  &  89.42 $\pm$ 1.68    &96.28 $\pm$ 2.82                 \\
         Appearance \& Spatial &\textit{AGS}& Random & 88.11 $\pm$ 2.47& 93.25 $\pm$ 1.94 \\
         Appearance \& Spatial &\textit{AGS} & Farthest &89.47 $\pm$ 2.71 &96.28 $\pm$ 1.03 \\
         Spatial& \textit{AGS}  & Fuse   & 69.50 $\pm$ 3.56&86.63 $\pm$ 4.67  \\
         Appearance& \textit{AGS}  & Fuse   &89.68  $\pm$ 2.28&\textbf{97.00 $\pm$ 1.10}  \\
         \toprule
         Appearance \& Spatial&\textit{AGS} & Fuse  &\textbf{91.60 $\pm$ 1.26}&\textbf{97.00 $\pm$ 1.10}  \\
         \toprule
    \end{tabular}
    \caption{Average patch-level accuracy and image-level accuracy on CRC dataset. GC represents the graph convolution method, where GS and AGS denote GraphSage and Adaptive GraphSage respectively. Sampling represents the nuclei sampling strategy.}
    \label{tab:big-table}
\end{table*}
\begin{figure*}[t]
\vspace{-15pt}
    \centering
    \includegraphics[width=1\textwidth]{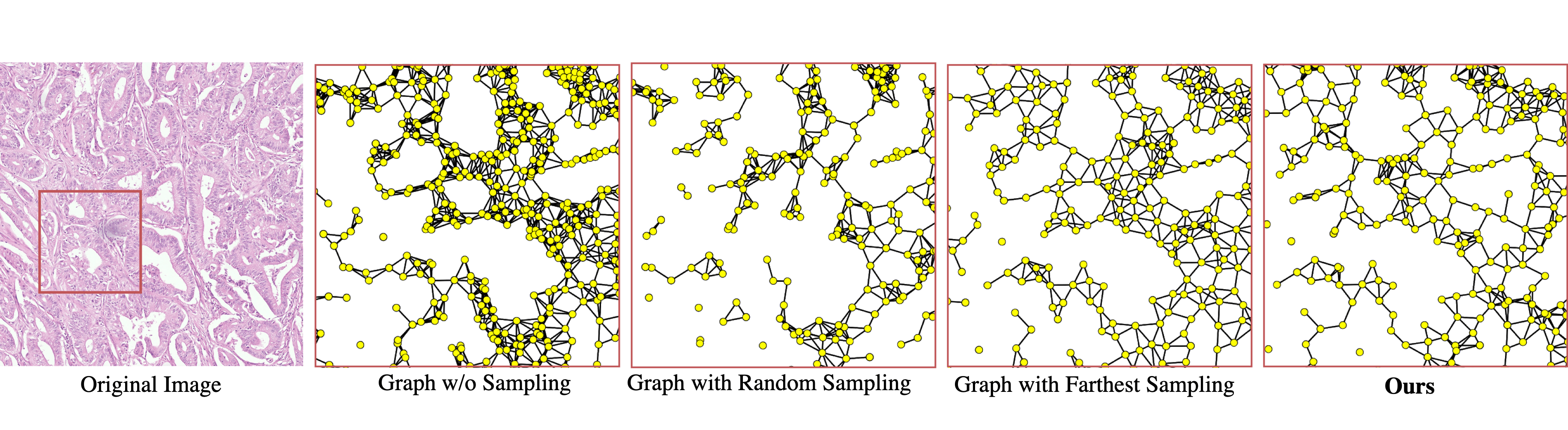}
     \vspace{-15pt}
    \caption{Comparison of different node sampling strategies.}
    \label{fig:sampling}
    \vspace{-10pt}
\end{figure*}
\begin{table}[h]
    \centering
\begin{tabular}{c|c}
\toprule
       Method &  Image Accuracy\\
       \toprule
       BAM-1 ~\cite{awan2017glandular} & 87.79 $\pm$ 2.32 \\
        BAM-2   ~\cite{awan2017glandular} & 90.66 $\pm$ 2.45\\
        Context - G  ~\cite{wsicontext2018} & 89.96 $\pm$ 3.54\\
        ResNet50 ~\cite{he2016deep}&92.08 $\pm$ 2.08  \\
        MobileNet  ~\cite{howard2017mobilenets}&  92.78 $\pm$ 2.74\\
        InceptionV3  ~\cite{szegedy2015going}& 91.37 $\pm$ 3.55 \\
        Xception ~\cite{chollet2017xception}&92.09 $\pm$ 0.98 \\
        CA-CNN  ~\cite{Shaban2019}& 95.70 $\pm$ 3.04\\
        \textbf{Ours} & \textbf{97.00 $\pm$ 1.10 } \\
    \toprule

\end{tabular}
    \caption{Comparison with state-of-the-art on CRC dataset.}
    \vspace{-15pt}
    \label{tab:compare-sota}
\end{table}

% \begin{table}[h]
%     \centering
% \begin{tabular}{c|c|c}
% \toprule
%       Method & Binary Acc. & Three-class Acc.\\
%       \toprule
%       BAM-1 & 95.70 $\pm$ 2.10 & 87.79 $\pm$ 2.32 \\
%         BAM-2    &97.12 $\pm$ 1.27 & 90.66 $\pm$ 2.45\\
%         Context - G   &96.44 $\pm$ 3.61 & 89.96 $\pm$ 3.54\\
%         ResNet50  & 98.57 $\pm$ 1.01&92.08 $\pm$ 2.08  \\
%         MobileNet &97.83 $\pm$ 1.77 &  92.78 $\pm$ 2.74\\
%         InceptionV3  & 98.57 $\pm$ 1.01& 91.37 $\pm$ 3.55 \\
%         Xception & 98.58 $\pm$ 2.01&92.09 $\pm$ 0.98 \\
%         CA-CNN& \textbf{99.28 $\pm$ 1.25}&  95.70 $\pm$ 3.04\\
%         \textbf{Ours} &98.52 $\pm$ 1.04  &  \textbf{97.00 $\pm$ 1.10 } \\
%     \toprule

% \end{tabular}
%     \caption{Comparison with state-of-the-arts on CRC dataset.}
%     \label{tab:compare-sota}
% \end{table}

\subsubsection{Comparison with State-of-the-Art}
To prove the effectiveness of CGC-Net, we compared the proposed method with recent state-of-the-art methods. \\
(1) \textbf{Context-Aware-CNN (CA-CNN)}~\cite{Shaban2019}: the state-of-the-art context-aware learning framework for CRC grading which incorporates large contextual regions and aggregates information via patch-based feature-cubes. \\
(2) \textbf{BAM}~\cite{awan2017glandular}: a two-step method for CRC grading which first segments glands and then computes the Best Alignment Metric (BAM) for classification. Here, we report the results of BAM-1 and BAM-2. BAM-1 computes the average BAM and the BAM entropy, whereas BAM-2 additionally computes the Regularity Index. \\
(3) \textbf{Context-G}~\cite{wsicontext2018}: A context-aware approach which makes use of a shared CNN followed by a long-short term memory (LSTM) to aggregate multi-scale information. \\
(4) Patch-based models: Several State-of-the-arts networks including ResNet50~\cite{he2016deep}, MobileNet~\cite{howard2017mobilenets}, InceptionV3~\cite{szegedy2015going} and Xception~\cite{chollet2017xception} are trained small patches of size 224$\times$224.
It must be noted that the 3 folds were split in the same way as ~\cite{Shaban2019} and therefore we utilise the comparative results from the respective paper.
\begin{figure*}[th]
    \centering
    \includegraphics[width=0.88\textwidth]{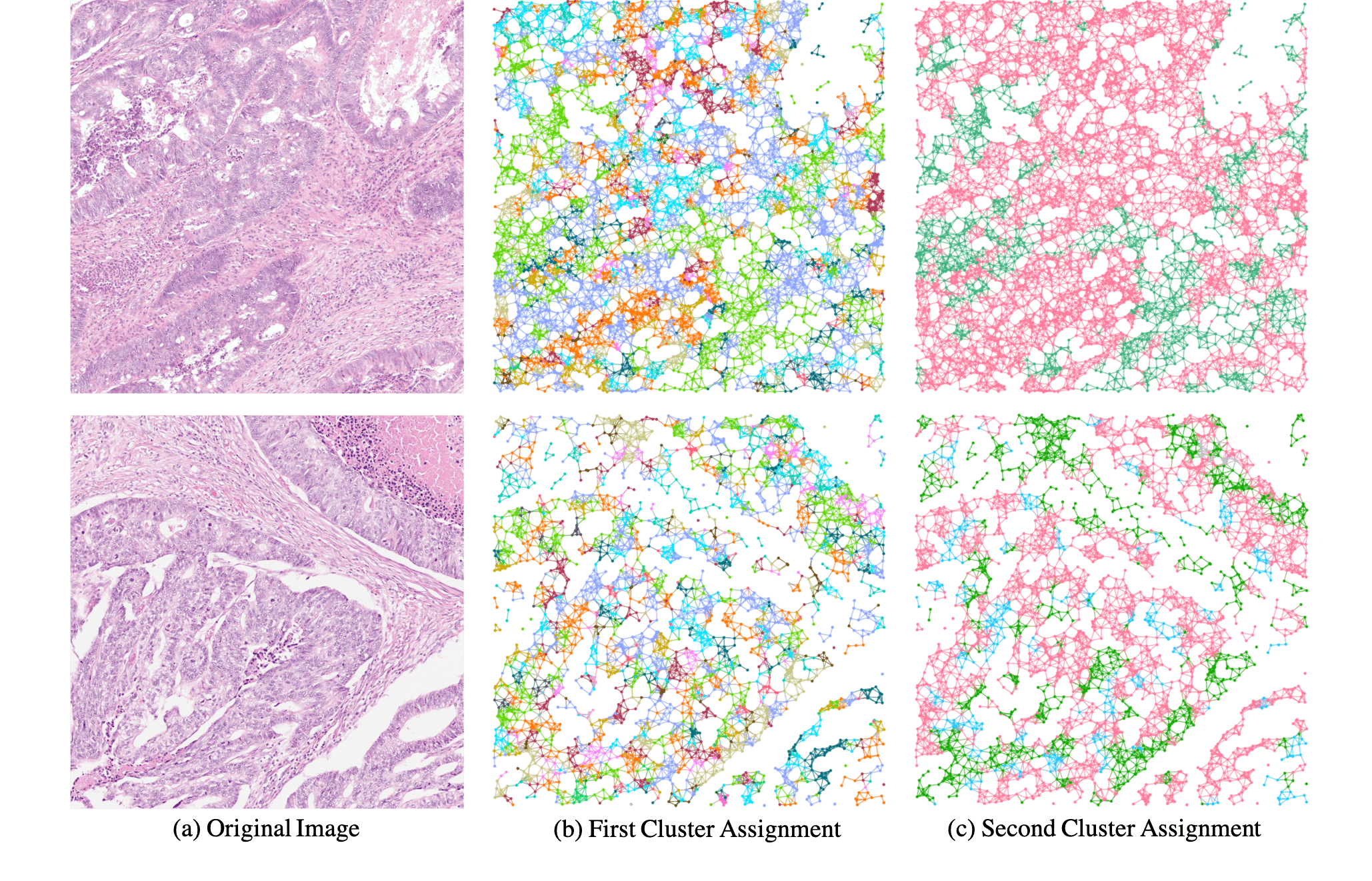}
    \caption{Visualisation of the node clustering results. Nodes with different colours represent that they are arranged to different clusters.}
    \vspace{-10pt}
    \label{fig:visual}
\end{figure*}
% \begin{table*}[t]
%     \centering
%     \begin{tabular}{c|c|c|c|c|c|c}
%     \toprule
%          Method   & Node Feature& GC       & Sampling& Patch Acc.& Three-class Acc.& Binary Acc. \\
%     \toprule
%          CG-Net-gs&  $app+coor$ &\textit{GS}& \textit{fuse}   &  89.42 $\pm$ 1.68    &96.28 $\pm$ 2.82    & 98.52 $\pm$ 1.05                 \\
%          CG-Net-random& $app+coor$&\textit{AGS}&\textit{random} & 88.11 $\pm$ 2.47& 93.25 $\pm$ 1.94&96.97 $\pm$ 2.84 \\
%          CG-Net-farthest&$app+coor$&\textit{AGS} & \textit{farthest}&89.47 $\pm$ 2.71 &96.28 $\pm$ 1.03 &99.24 $\pm$ 1.07 \\
%          CG-Net-c&$coor$& \textit{AGS}  & \textit{fuse}   & 69.50 $\pm$ 3.56&86.63 $\pm$ 4.67 &91.86 $\pm$ 4.45 \\
%           CG-Net-a&$app$& \textit{AGS}  & \textit{fuse}   &89.68  $\pm$ 2.28&97.00 $\pm$ 1.10 &98.48 $\pm$ 1.07 \\
%          \toprule
%          \textbf{Ours}&$app+coor$&\textit{AGS} &\textit{fuse}  &91.60 $\pm$ 1.26&97.00 $\pm$ 1.10 & 98.52 $\pm$ 1.05 \\
%          \toprule
%     \end{tabular}
%     \caption{Caption}
%     \label{tab:big-table}
% \end{table*}

As can be seen in Table~\ref{tab:compare-sota}, our proposed CGC-Net outperforms all competing methods by a large margin with smaller standard deviation, highlighting that CGC-Net is well suited to the task of colorectal cancer grading by incorporating both nuclear and graph-level features.
% \textbf{How to explain 2 class results?}
% Noticed that CG-Net does not achieve the best results for binary classification result.
% We hypothesize that contextual features aggregated from patches are strong enough to capture information from relative normal glandular tissues while the cell-graph has stronger capability to obtain information from \textit{high-grade} tissue with more irregular structure.

\subsubsection{Ablation Studies}
\noindent\textbf{Adaptive GraphSage:}
Inspired by GraphSage~\cite{hamilton2017inductive} and JK-Net~\cite{jknet}, we proposed a new graph convolution method for neighboring aggregation, named Adaptive GraphSage.
Compared with GraphSage~\cite{hamilton2017inductive}, the proposed method aggregates information within a neighbourhood adaptively, which is of significance for the cell graph of CRC with irregular glandular structure.
Comparative results are shown in Table~\ref{tab:big-table}, where we observe that our proposed method achieves the best performance on both patch-level and image-level accuracy.
\\
\noindent\textbf{Node Features:}
In order to test the effect of nucleus appearance features and spatial features, we construct the cell graph utilizing different features combination.
Results can be seen in Table~\ref{tab:big-table}.
Cell graph with only spatial features achieves $86.63\%$ for image-level accuracy, which demonstrates the tissue architecture is informative for CRC grading.
Meanwhile, the results of cell graph with nucleus appearance features are only $1.92\%$ below the cell graph with combined features on patch-level accuracy, which shows that the cellular heterogeneity in cancer region is a pivotal visual cue for cancer grading.
Graph equipped with combined features achieves the highest results, $91.60\%$ for patch accuracy and $97.00\%$ for image accuracy.
Though the image-level results are the same, utilizing complementary information from combined features has higher patch-level accuracy, which demostrates the patch predictions within one image are more consistent.
% Additionally, adding cell appearance features further boosts the performance to  $91.60\%$ for patch accuracy and $97.00\%$ for image accuracy, which shows the effectiveness of combing complementary information.
% It highlights the effectiveness of modeling nuclear features along with their cellular interactions
% Results can be seen in Tabel. ~\ref{tab:big-table}. Cell graph equipped with the combination of appearance and spatial features achieves the highest result, which gets $91.60\%$ for patch accuracy and $97.00\%$ for image accuracy.
% Meanwhile, although the image-level result from the cell graph with nucleus appearance features is as high as that from the cell graph with combined features, the patch level accuracy
% Meanwhile, the results of cell graph with nucleus appearance features are only $1.92\%$ below the cell graph with combined features on patch level accuracy, which proves that the cellular heterogeneity in cancer region is a pivotal visual cue for cancer grading.
% Last but not least, image-level accuracy from cell graph with only spatial features also achieves $86.63\%$.
% It demonstrates the effectiveness of the cell community architecture information for cancer grading, which further proves the effectiveness of our motive to construct the cell graph.
\\
\noindent\textbf{Nuclei sampling strategy:}
As introduced in Section~\ref{sec:graph-construct}, to reduce the size of the cell graph and preserve the cell architecture information in the graph, we propose a representative nuclei sampling strategy.
Experiments of utilizing different sample strategy can be seen in Table~\ref{tab:big-table}, where Random, Farthest and Fuse denote the random sampling, farthest sampling and our proposed sampling method.
Compared with results from random sampling in the second row, using farthest sampling achieves improvement from $88.11\%$ to $89.47\%$ for the patch accuracy, as well as the image accuracy from $93.25\%$ to $96.28\%$.
This is because the farthest sampling can preserve the glandular structures better.
In addition, the patch accuracy is further improved from $89.47\%$ to $91.60\%$ using our proposed sampling methods, which is because it either preserves the tissue structures or reflects the cell density to some extent, which can be considered as a trade-off between them.

Visualisation of different cell graphs constructed by different sampling methods can be seen in Figure~\ref{fig:sampling}.
Cells in the original graph are extremely dense with a large number of edges, which is neither feasible nor necessary for training.
Random sampling leads to a large number of disjoint sub-graphs in the sparse cell regions and therefore does no appropriately model the micro-environment well.
% makes the cell graph contain several local density regions as well as the disjoint sub-graphs in sparse cell area, which breaks the tissue architecture and is further harmful to parse message in the CGC-Net.
In addition, farthest sampling leads to a better tissue structure representation, especially in regions with very few cells, but does not reflect the cell density well, which could also be informative when grading the cancer.
Lastly, our method effectively captures the structures and the cell density to some extent, which can better model the tissue micro-environment.

\subsection{Cell Graph Visualisation}
To fully understand our method, we visualise the cluster assignment of each node in the first and second layers of our proposed CGC-Net.
Specifically, we assign each node to a cluster ID according to the maximum probability in the assignment matrix.
For the second cluster assignment output, we map the cluster ID back to the original nodes.
As can be seen in Figure~\ref{fig:visual}, the first clustering operation groups cells with similar appearance into many small clusters, which shows the nodes have aggregated local contextual information.
After the second clustering operation, we observe that fewer clusters are present within the graph and cells belonging to similar tissue structures tend to gather together.
Specifically, we observe that nuclei within potential tumour regions are assigned to the red cluster and nuclei in normal tissue regions are assigned to the green cluster.
Overall, it demonstrates the CGC-Net hierarchically aggregates information and captures both the local and the global features. Not only this, but the cluster visualisation gives us confidence that the features that the graph convolutional network are learning are biologically plausible.

%% file: conclusion.tex
In this paper, we propose a novel first-of-its-kind cell graph convolutional network for grading of colorectal cancer histology images, effectively aggregating information about cell morphology and tissue micro-architecture  through a stack of graph convolution and clustering operations.
Furthermore, we propose a general cell-graph construction pipeline with a representative nuclei sampling strategy to reduce computational redundancy.
The proposed CGC-Net effectively models the tissue micro-environment by considering the appearance features of the cell nuclei along with their local interactions, outperforming current state-of-the-art methods on a large-scale colorectal cancer grading dataset.